# Alternatives to Mobile Keypad Design: Improved Text Feed

Satish Narayana Srirama
Masters in Software Systems Engineering
RWTH Aachen, Germany
Hans-Böckler-Allee, 155/ Zi 101
52074 – Aachen, Germany
Snsrirama@yahoo.com

Mohammad Abdullah Al Faruque
Masters in Software Systems Engineering
RWTH Aachen, Germany
Alsenstr,3 D-52068,Aachen ,Germany
mafs33@yahoo.com

Mst Ayesha Siddika Munni
Computer Science and Engineering, IST Bangladesh National University
Dhaka, Bangladesh
munni_cse@yahoo.com

**ABSTRACT**

*In this paper we tried to focus on some of the problems with the mobile keypad and text entering in these devices, and tried to give some possible suggestions. We mainly took some of the basic Human Computer Interaction principles and some general issues into consideration.*

## 1. INTRODUCTION

Today the usage of mobiles has increased drastically and people are trying to use it for every service. People are using it for making phone calls, sending SMS text messages and some basic research is going on even to make it a Web Service provider. This makes the mobile essential thing in almost all of our daily needs. Still there are some basic problems in these devices and these things are to be corrected as early as possible, or else people would be so accustomed to these problems that it would be later very tough to change the views of people.

## 2. PROBLEMS DESCRIPTION

There are many problems with the mobile keypad like the buttons are small and hard to enter text. Today mobiles are so small that holding them in a hand and entering the details are too tough and cause fatigue. Basically here we are not concentrating on all these physical details but mainly want to concentrate on problems while entering text.

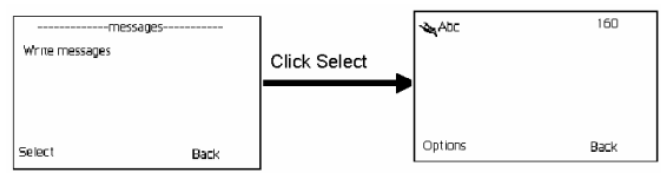

**Fig. 1: The change of the mobile screen during the message insertion phase**

Today SMS messaging and chatting have become basic needs for our daily needs. Say we are doing SMS chat and we have to enter messages. So for that in general we navigate through different menus and finally we reach the 'write message' screen then we select it. Ok leave this essential process, which can't be eliminated. In general the screen would appear as in Fig. 1.

**Fig. 2: The basic design of Mobile Keypad**

So now we are entering a text. We would press different buttons on the keypad multiple times to get the respective characters. Say for example to get character 'c' we press the button labelled 2 thrice, to get character '2' we press the button labelled 2 four times.

So while entering a text if we have to enter a number then the number of key selections would be more. But in general when we type say numbers, we do not stop with single numeric digit. If we are adding a number then our intension could be entering a phone no. Or some other numeric detail which could contain many numeric digits in succession. So in order to type a no like 02419964149. The no of keys selected would be 44. That counts to too much and in general sense we don't recognize this, as we are already accustomed to it.

Also the placement of the keys is not so perfect and can be upgraded. For example the normal Mobile keypad is designed taking that, the probability of getting any character in the text is the same. But in general the probability of occurrence of an oval is greater than the probability of occurrence of a consonant.

## 3. SUGGESTIONS FOR CHANGES IN DESIGN

So for the first problem discussed we could have a key, which should alter the mode of text between 'Alphabetic' and 'special chars' (Including numeric chars to special chars). It should be same as the 'Num-Lock' key on the normal Keyboard. There is no shortage for such a key as the ´#´ key is idle which can be used for this purpose. The Mid-fidelity design for the above problem is clearly shown in Fig. 3.

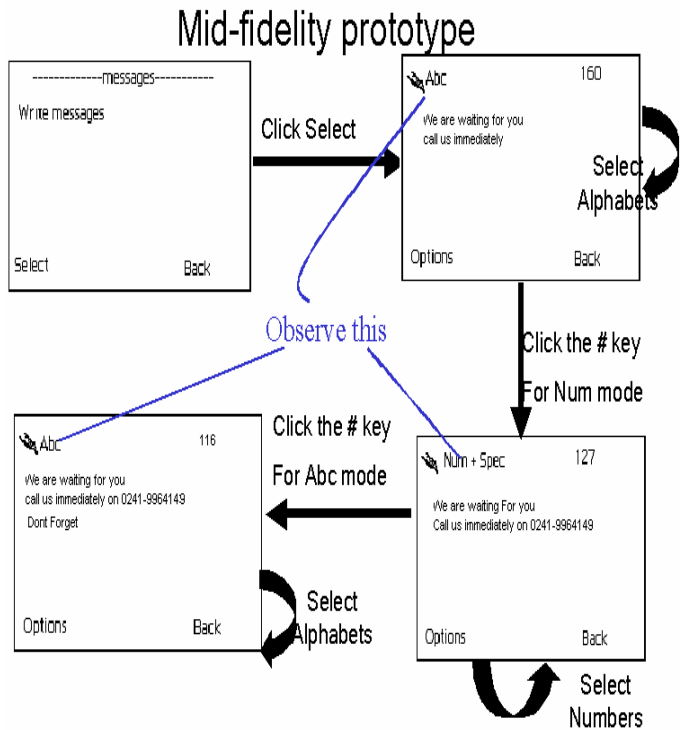

**Fig. 3: The mid-fidelity prototype of the design of the screen with the insertion of the change mode key in the keypad**

A solution for the second problem could really speed up the text processing by the user. The main basis for this paper is this problem and its solution. There could be many solutions. But we are providing this solution based on some of the HCI principles. We suggest a change in basic positioning of different keys. The solution design is in Fig. 4.

In general key pad the number of ovals at the first position is only one and by this modified design the Number of ovals at the first position is four. So this design could really speed up the process of entering text. (Basic research has to be done considering user and usability evaluation into account).
The Human Computer Interaction suggests that the design interfaces should not overload user's memories and they 'should promote recognition rather than recall'. So if we compare the basic design and the suggested design based on these principles, in the basic design the character are placed in their natural order so it is easy to recognize. Even in the suggested design the characters are placed in the proper cyclic order. So the load on the Cognition would be almost the same but the second design could really speed up the process.

| 1 | 2 abr | 3 cds |
|---|---|---|
| 4 eft | 5 ghu | 6 ijv |
| 7 klwx | 8 mny | 9 opqz |
| * | 0 | # |

**Fig. 4: The suggested design for the mobile keypad**

## 4. CONCLUSION

So, many changes have to be done to the basic text processing in mobile devices for better results and these designs and suggestions are to be implemented quickly because once people really get accustomed to it then changing basic designs could be very problematic. The general key board 'QWERTY' should be an eye-opener, as even though other keyboard designs allow faster typing, the large social base of QWERTY typists are reluctant to change.